\documentclass[aps,prb,twocolumn,amsmath,amssymb,groupedaddress,longbibliography
]{revtex4-2}

\usepackage{graphicx}
\usepackage{dcolumn}
\usepackage{bm}
\usepackage{multirow}
\usepackage{braket}
\usepackage{longtable}
\usepackage{color}
\usepackage{ulem}
\usepackage{booktabs}

\begin{document}

\title{
Nonlinear spin Hall effect in $\mathcal{PT}$-symmetric collinear magnets
}

\author{Satoru Hayami$^{1,2}$, Megumi Yatsushiro$^{1,2}$, and Hiroaki Kusunose$^3$}
\affiliation{
$^1$Department of Applied Physics, the University of Tokyo, Tokyo 113-8656, Japan \\
$^2$Department of Physics, Hokkaido University, Sapporo 060-0810, Japan \\
$^3$Department of Physics, Meiji University, Kawasaki 214-8571, Japan 
 }

\begin{abstract}
We theoretically investigate a nonlinear spin Hall effect in $\mathcal{PT}$-symmetric antiferromagnetic metals, which serve as an efficient spin current generator. 
We elucidate that an emergent spin-dependent Berry curvature dipole is a microscopic origin of the nonlinear spin Hall effect, which becomes nonzero with neither relativistic spin-orbit coupling, uniform magnetization, nor spin-split band structure. 
By analyzing a microscopic antiferromagnetic model without spin-orbit coupling for an intuitive understanding of the phenomena, we elucidate essential hopping processes and a condition to enhance the nonlinear spin Hall conductivity. 
We also provide a complete table to include useful correspondence among the N\'eel vector, odd-parity multipoles, nonlinear spin conductivity tensor, and candidate materials in all the $\mathcal{PT}$-symmetric black-and-white magnetic point groups.  
\end{abstract}

\maketitle

\section{Introduction}
The interplay between charge and spin degrees of freedom of electrons has long been studied in condensed matter physics, as it gives rise to rich physical phenomena, such as the anomalous Hall effect~\cite{Nagaosa_RevModPhys.82.1539,Xiao_RevModPhys.82.1959,nakatsuji2015large,Suzuki_PhysRevB.95.094406,vsmejkal2020crystal,Naka_PhysRevB.102.075112,Hayami_PhysRevB.103.L180407,smejkal2021anomalous} and magnetoelectric effect~\cite{Fiebig0022-3727-38-8-R01,Katsura_PhysRevLett.95.057205,Mostovoy_PhysRevLett.96.067601,tokura2014multiferroics}. 
Especially, the recent rapid development of controlling and detecting spin dynamics has opened up a field of spintronics~\cite{Zutic_RevModPhys.76.323,jungwirth2016antiferromagnetic, Baltz_RevModPhys.90.015005, Manchon_RevModPhys.91.035004}. 
The electric manipulation of spins as information carrier brings about the great potential for realizing high-speed, low-power, and high-capacity devices. 

One of the central issues in spintronics is to explore an efficient way of generating spin current. 
Although a metallic ferromagnetic system is a typical example to directly generate spin current~\cite{datta1990electronic, Gardelis_PhysRevB.60.7764, Schmidt_PhysRevB.62.R4790, Hu_PhysRevB.63.125333}, recent studies have clarified that noncentrosymmetric nonmagnetic systems with the relativistic spin-orbit coupling (SOC) are its good platform, since the latter is free from charge current and stray magnetic field. 
Indeed, a variety of mechanisms to induce pure spin current have been found so far, e.g., the spin Hall effect~\cite{d1971spin,hirsch1999spin,Zhang_PhysRevLett.85.393,murakami2003dissipationless,Sinova_PhysRevLett.92.126603,Murakami_PhysRevLett.93.156804,sinova2015spin},  circular photogalvanic effect ~\cite{Ganichev_PhysRevLett.86.4358,ganichev2002photogalvanic}, spin battery ~\cite{saitoh2006conversion,ando2011electrically,lesne2016highly,kondou2016fermi,Dushenko_PhysRevLett.116.166102}, and spin-rotation coupling effect~\cite{Matsuo_PhysRevB.87.180402,Kobayashi_PhysRevLett.119.077202}. 
In addition, nonlinear spin (Hall) current generation beyond the linear response regime has been formulated in noncentrosymmetric nonmagnetic systems with the SOC~\cite{Yu_PhysRevLett.113.156603, Hamamoto_PhysRevB.95.224430,araki2018strain, Pan_PhysRevB.99.245204, Zhang_PhysRevB.104.115140}. 
These SOC-based mechanisms have been successfully applied to a variety of materials. 

Meanwhile, another mechanism to generate spin current has been demonstrated in collinear antiferromagnets (AFMs) with neither the SOC nor uniform magnetization (stray magnetic field)~\cite{naka2019spin,hayami2019momentum, Yuan_PhysRevB.102.014422, Naka_PhysRevB.103.125114,Kondo_PhysRevResearch.4.013186}. 
The microscopic origin is attributed to an emergent $\bm{k}$-symmetric spin splitting in the band structure ($\bm{k}$ is the wave vector), which arises from an effective spin-dependent kinetic motion of electrons in AFM materials~\cite{Ahn_PhysRevB.99.184432,naka2019spin,hayami2019momentum,Yuan_PhysRevB.102.014422,Naka_PhysRevB.103.125114,shao2021spin,egorov2021colossal,egorov2021antiferromagnetism,Gonzalez_PhysRevLett.126.127701,bai2021observation}. 
Subsequently, the microscopic and macroscopic conditions in general AFMs inclduing noncollinear spin configurations have been clarified in terms of electronic-multipole description~\cite{hayami2019momentum,Hayami_PhysRevB.101.220403,Hayami2020b,Hayami_PhysRevB.102.144441}, perturbative approach ~\cite{Hayami_PhysRevB.105.024413}, and magnetic symmetry argument~\cite{Yuan_PhysRevB.102.014422,Yuan_PhysRevMaterials.5.014409,Yuan_PhysRevB.103.224410,vsmejkal2021altermagnetism}.  
These findings open up a new direction to explore further functional materials with negligibly small SOC, such as light-element materials, molecular organic metals, and 3$d$ transition metal oxides, out of conventional ones with strong SOC. 
In particular, collinear AFMs with negligibly small SOC are suitable for an efficient spin-current generator owing to their spin conservation.

\begin{figure}[t!]
\begin{center}
\includegraphics[width=1.0 \hsize ]{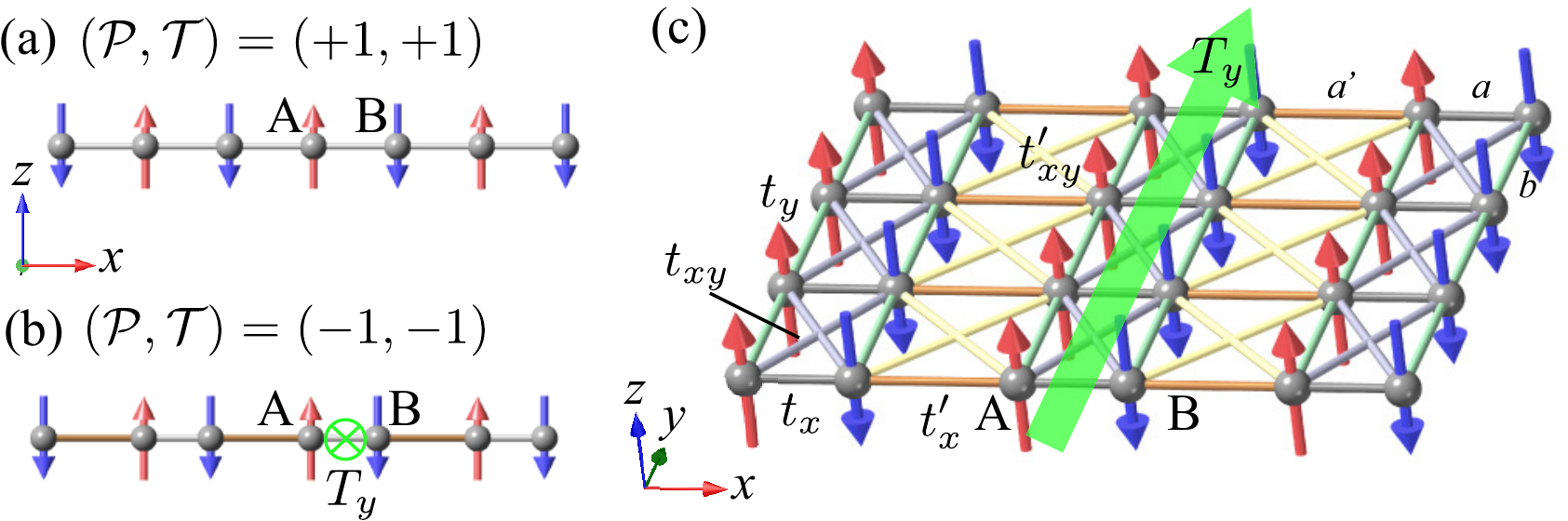} 
\caption{
\label{fig:fig1}
Staggered collinear $z$-spin alignment on the (a) uniform and (b) dimerized chains. 
(c) Collinear AFM on the two-dimensional orthorhombic system, where the hoppings in the model in Eq.~(\ref{eq:Ham}) are also shown. 
In (b) and (c), the AFM accompanies the \textit{uniform} magnetic toroidal dipole, $T_y$, along the $y$ direction. 
}
\end{center}
\end{figure}

In the present study, we propose a new type of spin current generation in collinear AFMs with $\mathcal{PT}$ symmetry via the nonlinear spin Hall effect, where $\mathcal{P}$ and $\mathcal{T}$ represent spatial inversion and time-reversal operations, respectively, which is qualitatively distinct from the linear spin-current generation from the symmetry viewpoint~\cite{naka2019spin,hayami2019momentum, Yuan_PhysRevB.102.014422, Naka_PhysRevB.103.125114}.
We find that a \textit{spin-dependent} Berry curvature dipole (BCD), which is induced by an effective spin-dependent hopping under AFM orderings, is a microscopic origin of a pure second-order spin current generation in metals without relying on the SOC, uniform magnetization, and any spin-split band structures. 
By analyzing a typical AFM system under $mm'm$ symmetry, we show that a large response is expected in the vicinity of the AFM phase transition, where the order parameter is small. 
Furthermore, we summarize general symmetry and AFM structures for nonzero nonlinear spin current conductivity under $\mathcal{PT}$-symmetric black-and-white magnetic point groups (MPGs) based on electronic odd-parity multipoles. 
Our results provide an alternative guideline to generate spin current in AFMs with neither the SOC nor spin-split band structures.

The rest of this paper is organized as follows. 
In Sec.~\ref{sec: Nonlinear spin Hall conductivity}, we show that the collinear magnets with a magnetic toroidal dipole moment lead to the nonlinear spin Hall conductivity. 
In Sec.~\ref{sec: Lattice model}, we demonstrate the emergence of the nonlinear spin Hall conductivity by considering the specific lattice model. 
We also discuss the important hopping parameters to induce it based on the real-space picture. 
Finally, we show the candidate materials to exhibit the nonlinear spin Hall conductivity under 16 black-and-white MPGs in Sec.~\ref{sec: Other magnetic point groups}. 
Section~\ref{sec: Summary} is devoted to a summary. 
In Appendix~\ref{sec: Nonlinear spin Hall effect with the spin-orbit coupling}, we show the influence of the SOC on the nonlinear spin Hall conductivity. 
In Appendix~\ref{sec: Irreducible representations of electric dipole and electric quadrupole}, we list the irreducible representations of electric
dipole and electric quadrupole to make the present paper self-contained. 

\section{Nonlinear spin Hall conductivity}
\label{sec: Nonlinear spin Hall conductivity}
Let us start with a two-sublattice AFM structure with $\mathcal{PT}$ symmetry in the one-dimensional chain in Figs.~\ref{fig:fig1}(a) and \ref{fig:fig1}(b); magnetic moments at sublattices $\zeta=$ A and B, $\bm{m}_{\zeta}$, satisfy $\bm{m}_{\rm A}=- \bm{m}_{\rm B}$. 
$\mathcal{P}$ and $\mathcal{T}$ symmetries are preserved on the uniform chain [Fig.~\ref{fig:fig1}(a)], while both are broken in the dimerized chain consisting of the staggered bond [Fig.~\ref{fig:fig1}(b)]. 
Such breakings of $\mathcal{P}$ and $\mathcal{T}$ symmetries lead to an emergent uniform magnetic toroidal dipole (MTD) $\bm{T}=(T_x,T_y,T_z)$ characterized by a $\mathcal{T}$-odd polar vector in the dimerized chain~\cite{Spaldin_0953-8984-20-43-434203,EdererPhysRevB.76.214404,Hayami_PhysRevB.90.024432,yatsushiro2021microscopic}. 
When the N\'eel vector, $\hat{\bm{h}}\parallel (\bm{m}_{\rm A}- \bm{m}_{\rm B})$, lies in the $z$ direction, the collinear AFM in Fig.~\ref{fig:fig1}(b) accommodates $T_y$ defined by the staggered magnetization $T_y \equiv m^z_{\rm A}- m^z_{\rm B}$~\cite{Hayami_PhysRevB.98.165110,Yatsushiro_PhysRevB.104.054412}, which is a source of nonlinear spin conductivity while vanishing linear one; no linear contribution is owing to the $\mathcal{PT}$-symmetric collinear AFM in the absence of the SOC. 
Hereafter, we consider the collinear AFM structure accompanying uniform $T_y$ in the two-dimensional system, where the dimerized chains along the $x$ direction are stacked in the $y$ direction as shown in Fig.~\ref{fig:fig1}(c). 
The situation is straightforwardly generalized to the cases with any N\'eel vectors under any point groups as shown in Table~\ref{table:mp}. 

\begin{figure}[t!]
\begin{center}
\includegraphics[width=1.0 \hsize ]{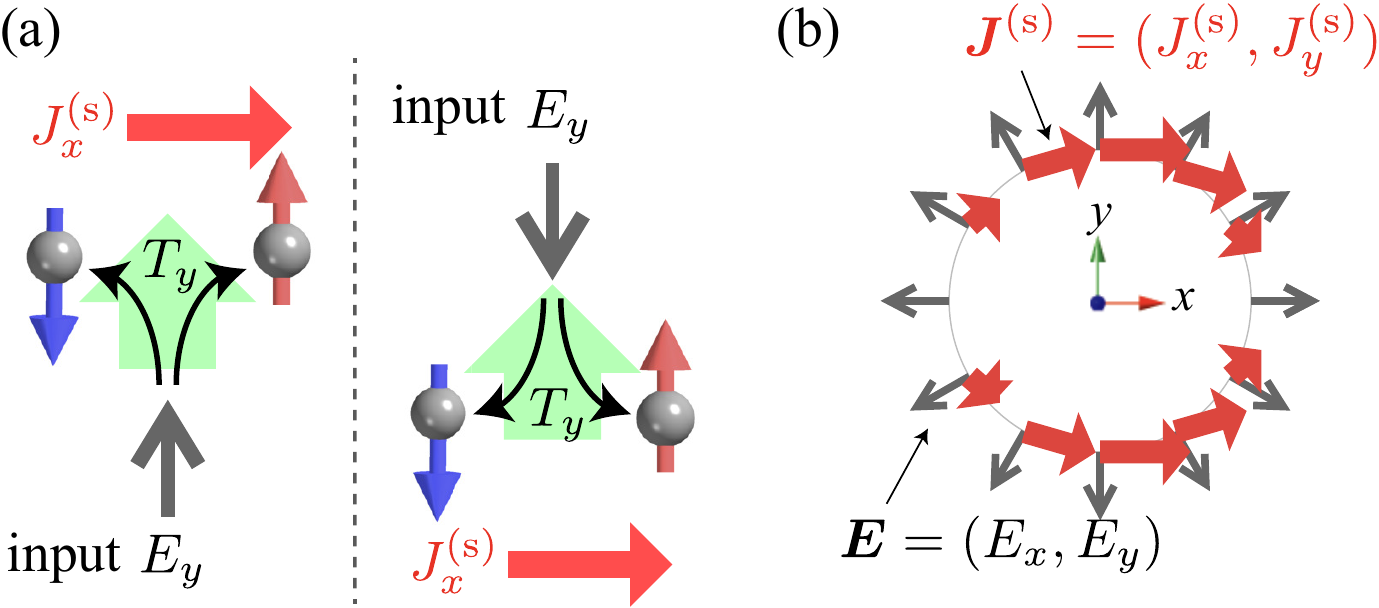} 
\caption{
\label{fig:fig2}
(a) Schematics of the nonlinear spin Hall effect when supposing $\sigma^{\rm (s)}_{x;yy}>0$ (the sign itself depends on material), where the spin current $J^{\rm (s)}_x$ is unchanged under the reversal of $E_y$. 
The green arrows represent the magnetic toroidal dipole $T_y$. 
(b) Field-angle dependence of the nonlinear spin current $\bm{J}^{\rm (s)}$ (red arrows) by the electric field $\bm{E}$ (gray arrows) when $\sigma^{\rm (s)}_{x;yy}>0$. 
}
\end{center}
\end{figure}

First, we show an intuitive picture of the nonlinear spin Hall conductivity $\sigma^{\rm (s)}_{\eta;\mu\nu}$ in $J^{\rm (s)}_{\eta}=\sigma^{\rm (s)}_{\eta;\mu\nu} E_{\mu}E_{\nu}$  in the presence of $\bm{T}$, where $J^{\rm (s)}_\eta$ and $E_{\mu,\nu}$ are a spin current and an electric field for the $\eta,\mu,\nu=x,y,z$ direction, respectively, and $\sigma^{\rm (s)}_{\eta;\mu\nu}=\sigma^{\rm (s)}_{\eta;\nu\mu}$. 
It is noted that the direction of the spin polarization in $J^{\rm (s)}_{\eta}$ is parallel to that of the N\'eel vector owing to the absence of the SOC; $J^{\rm (s)}_{\eta}= J_\eta \sigma$ with an electric current $J_\eta$ and spin $\sigma$ parallel to $\hat{\bm{h}}$. 
$\sigma^{\rm (s)}_{x;yy}$ has the same symmetry as $T_y$; both quantities are invariant under the symmetry operations of MPG $mm'm$.  
Thus, the resultant $J^{\rm (s)}_{x}$ is independent of $+E_y$ and $-E_y$, where the case of $\sigma^{\rm (s)}_{x;yy}>0$ is schematically shown in Fig.~\ref{fig:fig2}(a). 
Similarly, $\sigma^{\rm (s)}_{y;xy}$ also becomes nonzero under $T_y$. 

Microscopically, the nonlinear spin Hall conductivity is described by the spin-dependent BCD $D_n^{\mu\nu{\rm (s)}}(\bm{k})$, which is given by the second-order Kubo formula under the relaxation time approximation as~\cite{Sodemann_PhysRevLett.115.216806}
\begin{align}
\label{eq:BCD}
\sigma_{\eta;\mu\nu}^{\rm (s)}=\frac{e^3\tau}{2\hbar^2N} \sum_{\bm{k},n}f_{n\bm{k}}\epsilon_{\eta\mu\lambda}D_n^{\nu\lambda {\rm (s)}}(\bm{k}) + [\mu \leftrightarrow \nu], 
\end{align}
where $e$, $\tau$, $\hbar$, and $N$ are the electron charge, relaxation time, the reduced Planck constant, and the number of sites, respectively; we take $e=\tau=\hbar=1$. 
$f_{n\bm{k}}$ is the Fermi distribution function with the band index $n$ and $\epsilon_{\eta\mu\lambda}$ is the Levi-Civita tensor. 
In Eq.~(\ref{eq:BCD}), $D_n^{\mu\nu{\rm (s)}}(\bm{k})$ is related to the spin-dependent Berry curvature $\Omega_{n}^{\nu{\rm (s)}}(\bm{k})$ as $D_n^{\mu\nu{\rm (s)}}(\bm{k})=\partial_{\mu}\Omega_{n}^{\nu{\rm (s)}}(\bm{k})$~\cite{Sodemann_PhysRevLett.115.216806,Kondo_PhysRevResearch.4.013186}. 
Since we consider the collinear AFM structure, $D_n^{\mu\nu{\rm (s)}}(\bm{k})$ is expressed as the difference between up-spin and down-spin components similar to the spin Hall effect given by the spin-dependent Berry curvature~\cite{Kane_PhysRevLett.95.226801}. 
From the antisymmetric nature of $\sigma_{\eta;\mu\nu}^{\rm (s)}$ with respect to $\eta$ and $\mu$($\nu$), $\sigma^{\rm (s)}_{x;yy}$ is related to $\sigma^{\rm (s)}_{y;xy}$ as $2\sigma^{\rm (s)}_{x;yy}=-\sigma^{\rm (s)}_{y;xy}$; the factor 2 arises from the symmetrization regarding the input electric field.
It determines an angle dependence of the spin current generation against an electric field, as shown in Fig.~\ref{fig:fig2}(b).  
Remarkably, pure spin current occurs in the $x$ direction when the electric field is applied in the MTD ($y$) direction. 
Nonzero $\sigma_{\eta;\mu\nu}^{\rm (s)}$ requires both the $\mathcal{T}$- and $\mathcal{P}$-symmetry breaking, which indicates relevance to MTD and magnetic quadrupole (MQ), as detailed below, which is qualitatively different from the linear spin conductivity based on the magnetic toroidal quadrupole in the inversion-symmetric systems~\cite{hayami2022spinconductivity}.

\section{Analysis for the lattice model}
\label{sec: Lattice model}

To demonstrate the above consideration explicitly, we investigate the tight-binding model with $T_y$ as shown in Fig.~\ref{fig:fig1}(c), which is given by 
\begin{align}
\mathcal{H}= -\sum_{ij \sigma} 
t_{ij}
c_{i\sigma}^{\dagger}
c_{j\sigma}^{} - \sum_{i\sigma \sigma'} \bm{h}_i \cdot 
c_{i\sigma}^{\dagger} \bm{\sigma}_{\sigma \sigma'}c_{i\sigma'}^{},
\label{eq:Ham}
\end{align}
where $c^{\dagger}_{i\sigma}$ ($c_{i\sigma}^{}$) is the creation (annihilation) operator for site $i$ and spin $\sigma=\uparrow, \downarrow$. 
The first term represents the kinetic term; $t_{ij}$ includes the neighboring hoppings along the $x$ and $y$ directions, $(t_x, t_x', t_y)$, and the diagonal hoppings, $(t_{xy}, t'_{xy}$) in Fig.~\ref{fig:fig1}(c).
The second term represents the AFM mean-field term to satisfy $\bm{h}_{\rm A}=-\bm{h}_{\rm B}$ with the magnitude $|\bm{h}_{\rm A}|=h$ ($\bm{\sigma}$ is the vector of the Pauli matrices). 
We suppose the N\'eel vector along the $z$ direction as $\bm{h}_{\rm A}=(0,0,h)$ and $\bm{h}_{\rm B}=(0,0,-h)$ without loss of generality; there is no spin-split band structure owing to $\mathcal{PT}$ symmetry. 
In the following, we set $t_x=1$ as the energy unit, and set $t_y=0.7$, $t_{xy}=0.3$, $t'_x=\alpha t_x$, $t'_{xy}=\alpha t_{xy}$, and $\alpha=0.8$. 
We take the lattice constants as $a=a'=1$ and $b=1$ in Fig.~\ref{fig:fig1}(c).

\begin{figure}[t!]
\begin{center}
\includegraphics[width=1.0 \hsize ]{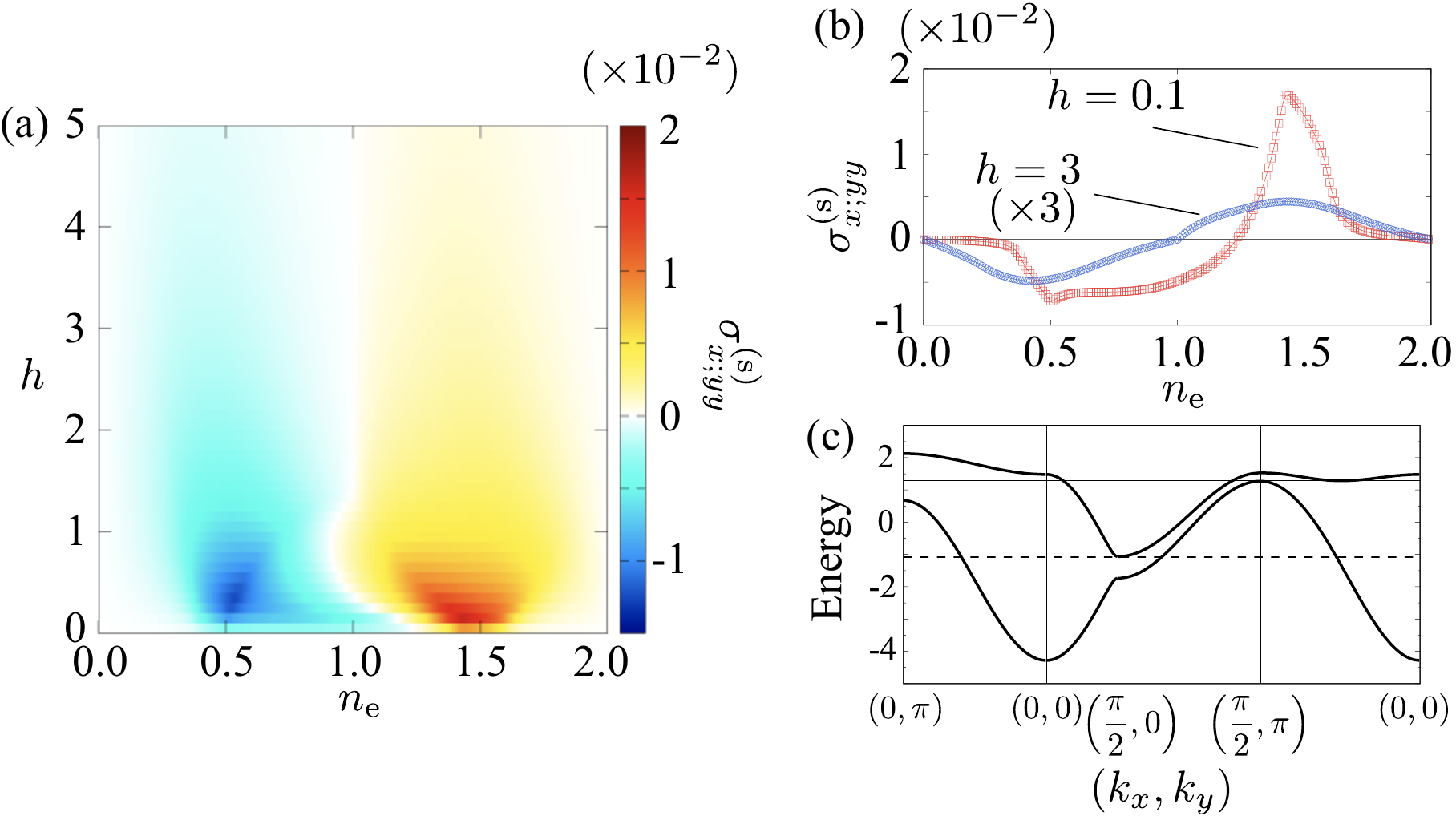} 
\caption{
\label{fig:fig3}
(a) Contour plots of $\sigma^{\rm (s)}_{x;yy}$ while changing $n_{\rm e}$ and $h$ at $t'_x=0.8$, $t_y=0.7$, $t_{xy}=0.3$, and $t'_{xy}=0.24$. 
(b) $n_{\rm e}$ dependence of $\sigma^{\rm (s)}_{x;yy}$ at $h=0.1$ and $h=3$. 
(c) The band structure at $h=0.1$. 
Each band is doubly degenerate owing to $\mathcal{PT}$ symmetry. 
The horizontal solid (dashed) lines in (c) represent the chemical potential that gives the largest (smallest) $\sigma^{\rm (s)}_{x;yy}$ in (b). 
}
\end{center}
\end{figure}

Figure~\ref{fig:fig3}(a) shows $\sigma^{\rm (s)}_{x;yy}$ calculated by Eq.~(\ref{eq:BCD}) while changing the electron density $n_{\rm e}$ ($n_{\rm e}=2$ corresponds to full filling) and $h$ at the temperature $T = 0.01$ and $N = 2\times 2400^2$. 
We also show $n_{\rm e}$ dependence of $\sigma^{\rm (s)}_{x;yy}$ at $h=0.1$ and $h=3$ in Fig.~\ref{fig:fig3}(b). 
The results clearly show nonzero $\sigma^{\rm (s)}_{x;yy}$ except for the insulating region at $n_{\rm e}=1$ and $h \gtrsim 1.6$ in contrast to the Zener-tunneling mechanism~\cite{Suzuki_PhysRevB.105.075201}. 
Especially, $|\sigma^{\rm (s)}_{x;yy}|$ is largely enhanced in the small $h$ region, as shown in Fig.~\ref{fig:fig3}(a), since its microscopic process in Eq.~(\ref{eq:BCD}) consists of the mutual interplay between the interband and intraband processes; both the small gap between the energy bands at the same $\bm{k}$ owing to small $h$ and large density of states is important to obtain large $|\sigma^{\rm (s)}_{x;yy}|$. 
Indeed, the maximum (minimum) value of $\sigma^{\rm (s)}_{x;yy}$ at $h=0.1$ is obtained when the chemical potential denoted as solid (dashed) lines lies on the band top (bottom) with a small separation from the other band so that the above conditions are satisfied, as shown in Fig.~\ref{fig:fig3}(c).
In such a situation, the dipole-like structure of the Berry curvature is remarkable; the monopole and anti-monopole in momentum space approach each other.
Meanwhile, in the large $h$ region where the upper and lower bands are well-separated, $\sigma^{\rm (s)}_{x;yy}$ is suppressed and shows an almost symmetric behavior against $n_{\rm e}$ in Fig.~\ref{fig:fig3}(b). 

The appearance of $\sigma^{\rm (s)}_{x;yy}$ depends on the type of the hopping parameters at the microscopic level. 
To obtain the essential model parameters for $\sigma^{\rm (s)}_{x;yy}$, we adopt an expansion method in the nonlinear conductivity tensor~\cite{Oiwa_doi:10.7566/JPSJ.91.014701}, where $\sigma^{\rm (s)}_{x;yy}$ is expanded as a polynomial form of products among the $i$th power of the Hamiltonian matrix at wave vector $\bm{k}$, $H^i(\bm{k})$, and the velocity operator, $\bm{v}_{\bm{k}}=\partial \mathcal{H}(\bm{k})/\partial \bm{k}$, as $\sigma_{x;yy}^{\rm (s)}=\sum_{ijk}C^{ijk}\sum_{\bm{k}}{\rm Tr}[v_{x \bm{k}}\sigma H^i(\bm{k})v_{y \bm{k}}H^j(\bm{k})v_{y \bm{k}}H^k(\bm{k})]$; $C^{ijk}$ is the model-independent coefficient.  
$\sigma^{\rm (s)}_{x;yy}$ is obtained as the imaginary part after taking the trace.

\begin{figure}[t!]
\begin{center}
\includegraphics[width=1.0 \hsize ]{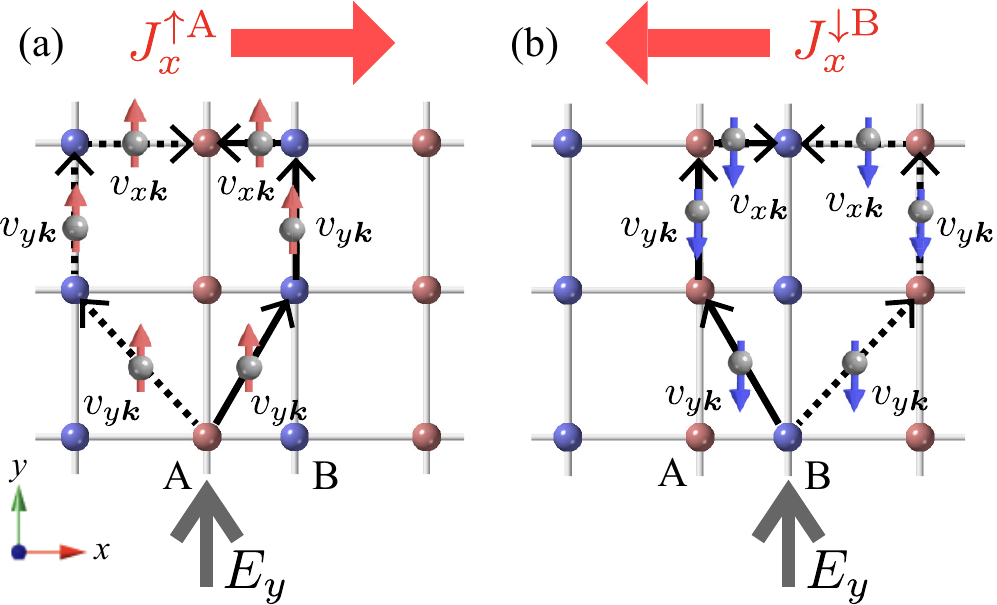} 
\caption{
\label{fig:fig4}
Hopping paths contributing to the spin current $J^{\sigma {\rm (\zeta)}}_x$ ($\sigma= \uparrow, \downarrow$ and $\zeta=$A and B) for the sublattices (a) A and (b) B under the AFM ordering with $E_y$. 
The inequivalence between the paths denoted by the solid and dashed arrows results in nonzero $J^{\sigma {\rm (\zeta)}}_x$.
The red and blue spheres represent up and down spins along the N\'eel vector, respectively. 
}
\end{center}
\end{figure}

The lowest-order contribution to $\sigma^{\rm (s)}_{x;yy}$ is given for $i=j=0$ and $k=1$ in the form of $h t_y (t_x t_{xy} - t'_x t'_{xy})$. 
In a similar way, the expansion includes various terms including the other hopping combinations, e.g., $h\left[2(t_{xy}^2-t'^2_{xy})F_1+(t_{x}^2 - t'^2_{x})F_2+ (t_{x} t_{xy}-t'_{x} t'_{xy}) F_3\right]$ in the second-lowest-order contribution for $i=0$ and $j=k=1$, where $F_1=t_y^2+t_{xy}^2+t'^2_{xy}$, $F_2=t_{xy}^2 + t'^2_{xy}$, and $F_3=t_{x} t_{xy}+t'_{x} t'_{xy}$. 
The important observation is that $\sigma^{\rm (s)}_{x;yy}$ becomes nonzero for 
$t_{xy} \neq t'_{xy}$ when $t_x=t'_x$ and $t_{xy} \neq 0$ or $t'_{xy} \neq 0$ when $t_x \neq t'_x$ in addition to $h \neq 0$. 
Hence, either $t_{xy}$ or $t'_{xy}$ in the diagonal hoppings is essential to give rise to $\sigma^{\rm (s)}_{x;yy}$. 

The importance of the diagonal hopping is understood from the matrix analysis in the trace. 
When considering the lowest-order process ($i=j=0$ and $k=1$), the contribution in the trace is simplified as ${\rm Tr}[v_{x \bm{k}}\sigma v_{y \bm{k}}v_{y \bm{k}}H(\bm{k})]= -h {\rm Tr}[v_{x \bm{k}}v_{y \bm{k}}v_{y \bm{k}} \rho_z]$ where $\bm{\rho}=(\rho_x,\rho_y,\rho_z)$ represents the vector of the Pauli matrices in A-B sublattice space. 
Thus, $\sigma^{\rm (s)}_{x;yy}$ remains finite only when an effective current $v_{x \bm{k}} v_{y \bm{k}}v_{y \bm{k}}$ has the same symmetry as $\rho_z$. 
As the $t_y$ ($t_x$, $t'_x$, $t_{xy}$, and $t'_{xy}$) term is proportional to $\rho_0$ ($c_1 \rho_x + c_2 \rho_y$ where $c_1$ and $c_2$ are $\bm{k}$-dependent constants), $t_{xy}$ or $t'_{xy}$ must be included in $v_{y\bm{k}}$ for ${\rm Tr}[v_{x \bm{k}}v_{y \bm{k}}v_{y \bm{k}} \rho_z] \neq 0$. 
Moreover, one also finds that $\sigma^{\rm (s)}_{x;yy}$ vanishes after taking the $k_x$ summation when $t_{xy}=t'_{xy}=0$.  
A similar argument holds for any higher-order contributions. 
In this way, the diagonal hopping is necessary for $\sigma^{\rm (s)}_{x;yy}$ in the present system.

\begin{table*}[t!]
\centering
\caption{
Conditions of the N\'eel vector $\hat{\bm{h}}$ to obtain nonlinear spin-current conductivity $\sigma^{\rm (s)}_{\eta;\mu\nu}$ in Eq.~(\ref{eq:BCD}) in the $\mathcal{PT}$-symmetric MPG except for $\bar{1}'$. 
See also Eq.~(\ref{eq:tensor}) for the symbols $(Q_x, Q_y, Q_z)$ and $(G_u, G_{v}, G_{yz}, G_{zx}, G_{xy})$. 
The active MTD $(T_x, T_y, T_z)$ and MQ $(M_u, M_{v}, M_{yz}, M_{zx}, M_{xy})$ and the candidate materials are also shown. 
In $4'/m'$, $4/m'$, $6/m'$, and $\bar{3}'$, $\hat{\bm{h}}\parallel \hat{\bm{x}}$ and $\hat{\bm{h}}\parallel \hat{\bm{y}}$ are common. 
\label{table:mp}}
\begin{tabular}{llllll}\hline \hline
MPG&  $\hat{\bm{h}} \parallel \hat{\bm{x}}$ & $\hat{\bm{h}} \parallel \hat{\bm{y}}$ & $\hat{\bm{h}} \parallel \hat{\bm{z}}$  & MTD, MQ& Materials  \\\hline
 
$4/m'm'm'$ 
& $Q_x, G_{yz}$ & $Q_y, G_{zx}$ & $Q_z$
&
$M_u$ & Fe$_2$TeO$_6$~\cite{kunnmann1968magnetic}, UBi$_2$~\cite{leciejewicz1967neutron}, UGeSe~\cite{ptasiewicz1978magnetic}, UPt$_2$Si$_2$~\cite{steeman1990hybridisation,amitsuka1992high}\\

$4'/m'm'm$ 
&$Q_x, G_{yz}$ & $Q_y, G_{zx}$ & $G_{xy}$
& 
$M_{v}$ & BaMn$_2$(As,Bi,P,Sb)$_2$~\cite{Singh_PhysRevB.80.100403,Calder_PhysRevB.89.064417,brock1994resistivity,Lovesey_PhysRevB.98.054434,Zhang_PhysRevB.99.184416,Huynh_PhysRevB.99.195111}, EuMnBi$_2$~\cite{Masuda_PhysRevB.101.174411}, CoAl$_2$O$_4$~\cite{Roy_PhysRevB.88.174415} \\
&&&&& (Ca,Sr)MnBi$_2$~\cite{Guo_PhysRevB.90.075120}, Ce$_2$PdGe$_3$~\cite{Bhattacharyya_PhysRevB.94.014418}, CeMnAsO~\cite{Zhang_PhysRevB.91.064418}  \\

 $4/m'mm$ 
&$Q_y, G_{zx}$ & $Q_x, G_{yz}$ & $G_{u}$
& 
$T_z$ & Co$_3$Al$_2$Si$_3$O$_{12}$~\cite{Cui_PhysRevB.101.144424}\footnote{Noncollinear spin textures have been reported.}  \\

$4'/m'$ 
&$Q_x, G_{yz}$ & $Q_y, G_{zx}$ & $G_{v}, G_{xy}$
&
$M_{v}$, $M_{xy}$ 
& K(Os,Ru)O$_4$~\cite{marjerrison2016structure,Hayami_PhysRevB.97.024414,injac2019structural,Yamaura_PhysRevB.99.155113}\\

$4/m'$ 
&$Q_x, G_{yz}$&$Q_y, G_{zx}$  &$Q_z, G_u$
& 
$T_z$, $M_u$ 
& (K,Rb)$_y$Fe$_{2-x}$Se$_2$~\cite{pomjakushin2011room}, TlFe$_{1.6}$Se$_2$~\cite{May_PhysRevLett.109.077003}, K$_{0.8}$Fe$_{1.8}$Se$_2$~\cite{wei2011novel} \\
\hline

$mmm'$ 
& $Q_y, G_{zx}$ & $Q_x, G_{yz}$ & $G_{u}, G_{v}$
& 
$T_z$, $M_{xy}$ 
& CuMnAs~\cite{wadley2013tetragonal,wadley2016electrical}, Mn$_2$Au~\cite{barthem2013revealing,bodnar2018writing}, 
Cr$_2$(W,Te)O$_6$~\cite{kunnmann1968magnetic,Zhu_PhysRevLett.113.076406},\\ 
&&&&& CeCoSi~\cite{lengyel2013temperature,tanida2018substitution,nikitin2020gradual,yatsushiro2020odd,Yatsushiro_PhysRevB.102.195147}, MnPd$_2$~\cite{kadar1972new}, Gd$_5$Ge$_4$~\cite{Tan_PhysRevB.71.214408},  
(Dy,Er)B$_4$~\cite{will1979neutron,will1981neutron} \\

$m'm'm'$ 
&$Q_x, G_{yz}$& $Q_y, G_{zx}$& $Q_z, G_{xy}$
&  
$M_u$, $M_{v}$ 
& LiMnPO$_4$~\cite{Toft_PhysRevB.85.224415}, GdAlO$_3$~\cite{quezel1982neutron}, EuMnSb$_2$~\cite{Soh_PhysRevB.100.174406,Gong_PhysRevB.101.224422}, TbB$_4$~\cite{will1981neutron}\\
\hline

$2'/m$ 
&$Q_y, G_{u}$ & $Q_{x}, Q_{z}$ & $Q_y, G_{u}$
& 
$T_z$, $T_x$ 
& CaMnSb$_2$~\cite{ratcliff2009magnetic}, MnPS$_3$~\cite{Ressouche_PhysRevB.82.100408}, MnGeO$_3$~\cite{redhammer2011magnetic}, LiCoPO$_4$~\cite{Vaknin_PhysRevB.65.224414} \\
 &
$G_v, G_{zx}$
&
$G_{yz}, G_{xy}$
&
$G_v, G_{zx}$
&$M_{xy}$, $M_{yz}$
& YbCl$_3$~\cite{Xing_PhysRevB.102.014427}, SrMn$_2$As$_2$~\cite{das2016collinear}, (K,Rb)FeS$_2$~\cite{bronger1987antiferromagnetic} \\

$2/m'$ 
&$Q_{x}, Q_{z}$ &$Q_y, G_{u}$ & $Q_{x}, Q_{z}$
& 
$T_y$, $M_u$ 
& LiFePO$_4$~\cite{Toft_PhysRevB.92.024404}, (Co, Fe)$_4$Nb$_2$O$_9$~\cite{Khanh_PhysRevB.93.075117,Khanh_PhysRevB.96.094434,Deng_PhysRevB.97.085154,Yanagi_PhysRevB.97.020404,Ding_PhysRevB.102.174443}, ErGe$_3$~\cite{schobinger1996magnetic} \\
&
$G_{yz}, G_{xy}$
&
$G_v, G_{zx}$
&
$G_{yz}, G_{xy}$
& $M_v$, $M_{zx}$ 
& CaMnGe~\cite{welter1996neutron}, KFeSe$_2$~\cite{bronger1987antiferromagnetic}, Fe$_2$Co$_2$Nb$_2$O$_9$~\cite{maignan2021fe} \\
\hline

$6/m'm'm'$ 
&$Q_x, G_{yz}$ & $Q_y, G_{zx}$ &$Q_z$
& 
$M_u$ \\

$6/m'mm$ 
&$Q_y, G_{zx}$ & $Q_x, G_{yz}$ & $G_u$
& 
$T_z$
\\

$6/m'$ 
&$Q_x, G_{yz}$ & $Q_y, G_{zx}$& $Q_z, G_u$
&  
$T_z$, $M_u$ 
& U$_{14}$Al$_{51}$~\cite{brown1997structure}\footnotemark[1] \\
\hline

$\bar{3} 'm'$ 
&$Q_x, G_{yz}$& $Q_y, G_{zx}$&$Q_z$
& 
$M_u$ 
& Cr$_2$O$_3$~\cite{McGuire_PhysRev.102.1000,brown2002determination}, Mn$_4$Ta$_2$O$_9$~\cite{Narayanan_PhysRevB.98.134438}, AgRuO$_3$~\cite{Schnelle_PhysRevB.103.214413} \\

$\bar{3}'m$ 
&$Q_y, G_{zx}$&$Q_x, G_{yz}$&$G_{u}$
& 
$T_z$ 
& Ca$_2$(Y,La)Zr$_2$Fe$_3$O$_{12}$~\cite{milam2021influence}\footnotemark[1] \\

$\bar{3}'$ 
&$Q_x, G_{yz}$ &$Q_y, G_{zx}$&$Q_z$, $G_u$
& 
$T_z$, $M_u$ 
& MgMnO$_3$~\cite{Haraguchi_PhysRevMaterials.3.124406} \\  
  
\hline\hline
\end{tabular}
\end{table*}

We further discuss the real-space hopping processes to get a more intuitive physical understanding in Fig.~\ref{fig:fig4}. 
In Figs.~\ref{fig:fig4}(a) and \ref{fig:fig4}(b), we exhibit representative lowest-order hopping processes contributing to $\sigma^{\rm (s)}_{x;yy}$ for the sublattice A and B, respectively. 
Owing to the $\mathcal{PT}$-symmetric AFM mean field, the role of the electrons with up spin ($\uparrow$) at the A sublattice is equivalent to that of down spin ($\downarrow$) at the B sublattice from the symmetry point of view. 
For the A sublattice in Fig.~\ref{fig:fig4}(a), the processes denoted as solid arrows and dashed arrows give different contributions proportional to $(t_x t_{xy} - t'_x t'_{xy})$, which results in the local up-spin current in the $x$ direction $J^{\uparrow {\rm (A)}}_x$. 
Similarly, for the B sublattice in Fig.~\ref{fig:fig4}(b), the local down-spin current with the same magnitude but the opposite direction, i.e., $J^{\downarrow {\rm (B)}}_x=-J^{\uparrow {\rm (A)}}_x$, occurs. 
Thus, pure spin current without charge current is generated in the $x$ direction.

Let us briefly discuss the influence of the SOC on $\sigma^{\rm (s)}_{x;yy}$. 
By calculating $\sigma^{\rm (s)}_{x;yy}$ for the model in Eq.~(\ref{eq:Ham}) with the SOC, we found that the contribution from the SOC is comparable to that from the present mechanism for weak $h$, while there is almost no influence of the SOC for large $h$, as detailed in Appendix~\ref{sec: Irreducible representations of electric dipole and electric quadrupole}. 
In other words, the present mechanism is important even when considering the SOC.
This is because our mechanism depends on the magnitudes of the order parameter of the magnetism and hopping integrals, which are usually much larger than the SOCs.

\section{Other magnetic point groups}
\label{sec: Other magnetic point groups}

Finally, we generalize our discussion to any types of $\mathcal{PT}$-symmetric collinear AFMs. 
In the presence of $\mathcal{PT}$ symmetry, $\sigma^{\rm (s)}_{\eta;\mu\nu}$ in Eq.~(\ref{eq:BCD}) can be finite when any of multipoles out of three MTDs $(T_x, T_y, T_z)$ and five MQs $(M_u, M_v, M_{yz}, M_{zx}, M_{xy})$ are activated under the AFM orderings.  
Namely, this situation is provided in 16 black-and-white MPGs~\cite{Yatsushiro_PhysRevB.104.054412}. 
We list the correspondence between MPGs and multipoles in Table~\ref{table:mp}. 
Nonzero components of $\sigma^{\rm (s)}_{\eta;\mu\nu}$ depend on the direction of the N\'eel vector $\hat{\bm{h}}$ and types of associated active multipoles. 
To extract such a dependence, we decompose $\sigma^{\rm (s)}_{\eta;\mu\nu}$ as $\sigma^{\rm (s)}_{\eta;\mu\nu}=\gamma_{\eta;\mu\nu}\sigma$, where $\gamma_{\eta;\mu\nu}$ is represented for $E_{\mu}E_{\nu} \to (E_x^2, E^2_y, E^2_z, E_{y} E_{z}, E_z E_x, E_x E_y)$ as 
\begin{align}
\label{eq:tensor}
\gamma_{\eta;\mu\nu} \propto
\left(
\begin{array}{ccc}
0 & 2(Q_{y}-G_{zx}) & 2(Q_z+G_{xy}) \\
2(Q_x+G_{yz}) & 0 & 2(Q_z-G_{xy}) \\
2(Q_x-G_{yz}) & 2(Q_{y}+G_{zx}) & 0 \\
G_u + G_v & -Q_{z}+G_{xy} & -Q_{y}-G_{zx} \\
-Q_{z}-G_{xy} & -G_u +G_v & -Q_{x}+G_{yz} \\
-Q_{y}+G_{zx} & -Q_{x}-G_{yz} & -2 G_v
\end{array}
\right)^{\rm T},
\end{align}
where ${\rm T}$ denotes the transpose of a matrix. 
Here, $(Q_x, Q_y, Q_z)$ and $(G_u, G_{v}, G_{yz}, G_{zx}, G_{xy})$ symbolically represent the $\mathcal{T}$-even polar vector with the same symmetry property as electric dipole and $\mathcal{T}$-even axial second-rank tensor components as electric toroidal quadrupole, respectively~\cite{Hayami_PhysRevB.98.165110,Yatsushiro_PhysRevB.104.054412}; their irreducible representation is listed in Appendix~\ref{sec: Irreducible representations of electric dipole and electric quadrupole}. 

Combining Table~\ref{table:mp} and Eq.~(\ref{eq:tensor}) provides information regarding nonzero $\sigma^{\rm (s)}_{\eta;\mu\nu}$. 
For example, when considering the $4/m'm'm'$ system with $\hat{h}_x$ ($\hat{h}_z$), one obtains $2\sigma^{\rm (s)}_{x;yy}=-\sigma^{\rm (s)}_{y;xy}$ and $2\sigma^{\rm (s)}_{x;zz}=-\sigma^{\rm (s)}_{z;zx}$ ($2\sigma^{\rm (s)}_{z;xx}=2\sigma^{\rm (s)}_{z;yy}=-\sigma^{\rm (s)}_{x;zx}=-\sigma^{\rm (s)}_{y;yz}$). 
In addition, it is noteworthy that the direction of the N\'eel vector $\hat{\bm{h}}$ is independent of tensor components. 
For instance, by changing the N\'eel vector from $\hat{\bm{h}}=(0,0,1)$ to $\hat{\bm{h}}=(1,0,0)$ in Fig.~\ref{fig:fig1}(c), the system symmetry changes from $mm'm$ to $m'm'm'$. 
Then, by looking at the column $\hat{\bm{h}} \parallel \hat{\bm{x}}$ and the row $m'm'm'$ in Table~\ref{table:mp}, one finds that the same multipoles, $Q_x$ and $G_{yz}$, contribute to $\sigma^{\rm (s)}_{\eta;\mu\nu}$, which results in the same tensor in the case of $\hat{\bm{h}}=(0,0,1)$. 
We list up the candidate materials to obtain finite $\sigma^{\rm (s)}_{\eta;\mu\nu}$ in accordance with MAGNDATA~\cite{gallego2016magndata}, magnetic structures database, and other references~\cite{schmid1973magnetoelectric,Watanabe_PhysRevB.98.245129}.
Although some materials in Table~\ref{table:mp} are categorized into insulators, nonzero $\sigma^{\rm (s)}_{\eta;\mu\nu}$ is still expected with a small insulating gap and/or by doping carriers. 
In addition, although the materials in Table~\ref{table:mp} have a considerable magnitude of the SOC, such an influence on $\sigma^{\rm (s)}_{\eta;\mu\nu}$ is expected to be negligibly small when the spontaneous staggered magnetization is well developed, as discussed above. 
Furthermore, as the symmetry conditions for the linear and nonlinear spin conductivity are different from each other~\cite{PhysRevB.92.155138}, one finds the situation where the pure nonlinear spin Hall conductivity driven by antiferromagnetic orderings can be observed depending on the direction. 
In particular, no linear contribution is expected in the $\mathcal{PT}$-symmetric collinear AFM in the absence of the SOC, as demonstrated in Sec.~\ref{sec: Lattice model}.

\section{Summary}
\label{sec: Summary}

To summarize, we have proposed spin current generation via the nonlinear spin Hall effect in $\mathcal{PT}$-symmetric collinear magnets. 
We clarified that its microscopic essence lies in nonzero spin-dependent Berry curvature dipole arising from the AFM phase transition accompanying the uniform magnetic toroidal dipole. 
As the present mechanism does not require any of the SOC, uniform magnetization, and any spin-split band structures, our results significantly broaden the scope of potential materials for next-generation spintronic devices based on AFMs. 
To stimulate further exploration of such functional materials, we provide a comprehensive table about the symmetry and AFM conditions for nonzero nonlinear spin current conductivity by using the concept of electronic odd-parity multipoles. 

\appendix
\section{Nonlinear spin Hall effect with the spin-orbit coupling}
\label{sec: Nonlinear spin Hall effect with the spin-orbit coupling}

In this Appendix, we investigate the influence of the SOC on the nonlinear spin Hall conductivity $\sigma^{\rm (s)}_{x;yy}$. 
For that purpose, we additionally consider the SOC Hamiltonian allowed from the symmetry of the lattice structure in Fig.~1 (c), which is represented by 
\begin{align}
\mathcal{H}^{\rm SOC}= \alpha^{\rm SOC} \sum_{\bm{k},\sigma,\sigma'}\sin k_y \sigma^z_{\sigma\sigma'} (
c^{\dagger}_{\bm{k}{\rm A}\sigma} c^{}_{\bm{k}{\rm A}\sigma'}
-c^{\dagger}_{\bm{k}{\rm B}\sigma}c^{}_{\bm{k}{\rm B}\sigma'}
), 
\end{align}
where $c_{\bm{k}\zeta\sigma}$ is the Fourier transform of $c_{i\sigma}$ with wave vector $\bm{k}$, sublattice $\zeta$, and spin $\sigma$.

\begin{figure}[h!]
\begin{center}
\includegraphics[width=0.8 \hsize ]{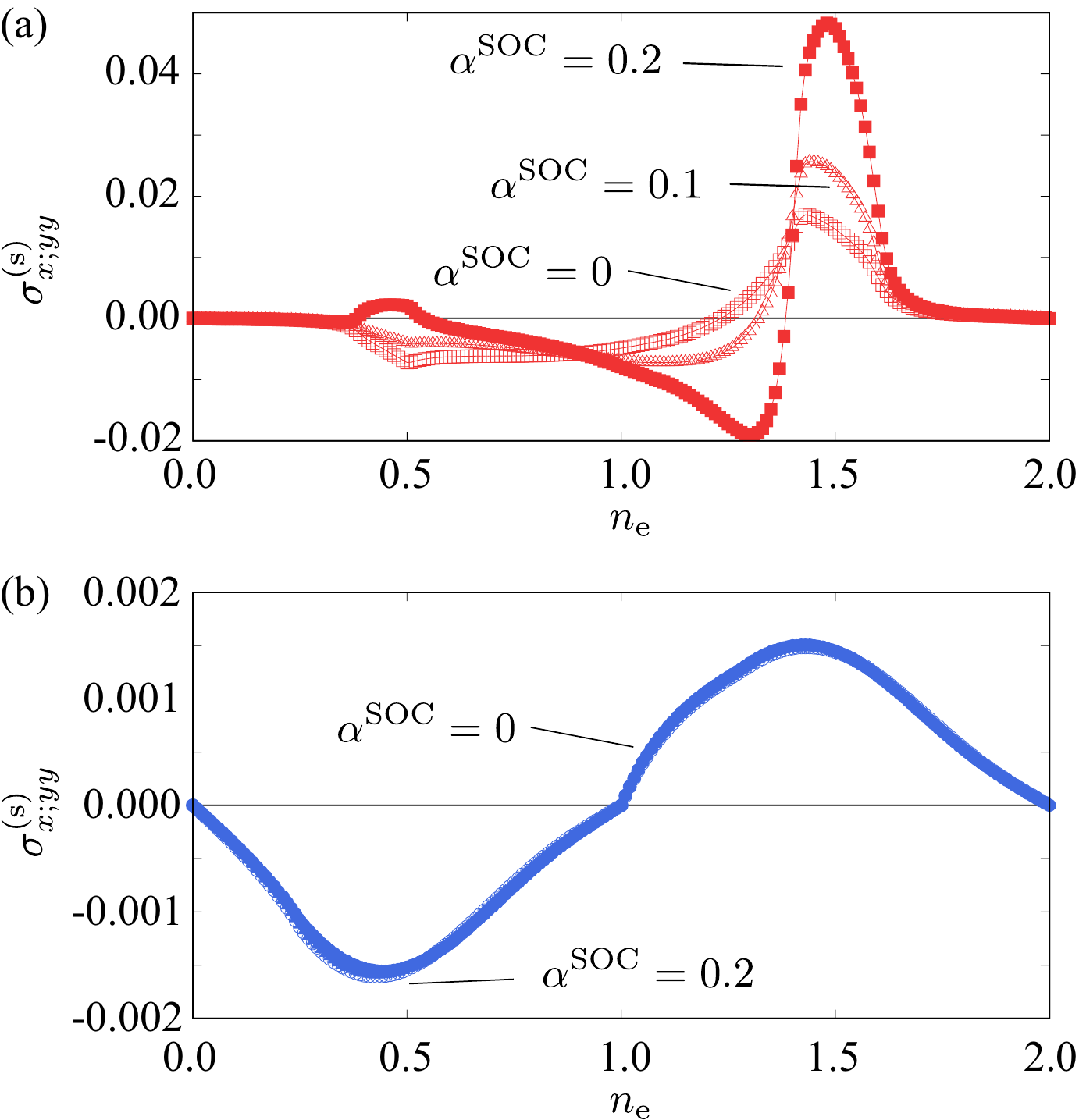} 
\caption{
\label{fig:SM}
$n_{\rm e}$ dependence of $\sigma^{\rm (s)}_{x;yy}$ in the presence of the SOC at (a) $h=0.1$ and (b) $h=3$. 
The SOC paramteres are taken at $\alpha^{\rm SOC}=0.1$ and $0.2$, where the data at $\alpha^{\rm SOC}=0.1$ in (b) is almost overlapperd with those at $\alpha_{\rm SOC}=0$.
}
\end{center}
\end{figure}

Figures~\ref{fig:SM}(a) and \ref{fig:SM}(b) show the $n_{\rm e}$ dependence of $\sigma^{\rm (s)}_{x;yy}$ for several values of $\alpha_{\rm SOC}=0.1, 0.2$ in the case of the small mean field $h=0.1$ and large meand field $h=3$, respectively. 
As a result, we found that the contribution of the SOC depends on the electron filling for small $h$, but their magnitude is in the same order as that from the result at $\alpha_{\rm SOC}=0$, as shown in Fig. S1(a). 
Furthermore, we found that there is almost no influence of the SOC for large $h$, as shown in Fig. S1(b). 
These results indicate that the present mechanism is important even when considering the SOC.

\section{Irreducible representations of electric dipole and electric quadrupole}
\label{sec: Irreducible representations of electric dipole and electric quadrupole}

As discussed in the main text, the rank-1 polar-vector quantity and rank-2 axial-tensor quantity have the same symmetry property as the electric dipole $(Q_x, Q_y, Q_z)$ and the electric toroidal quadrupole $(G_u, G_v, G_{yz}, G_{zx}, G_{xy})$, respectively. 
In this Appendix, we show the correspondence between these multipoles and their irreducible representation under cubic, tetragonal, orthorhombic, and monoclinic crystals in Table~\ref{tab_multipoles_table1} and under hexagonal and trigonal crystals in Table~\ref{tab_multipoles_table2}~\cite{Hayami_PhysRevB.98.165110}. 
Similarly, the irreducible representations in magnetic point groups are given by considering their unitary subgroup and antiunitary operation~\cite{Yatsushiro_PhysRevB.104.054412}.
The irreducible representation of $(Q_x, Q_y, Q_z)$ and $(G_u, G_v, G_{yz}, G_{zx}, G_{xy})$ gives information about the nonzero components of the nonlinear spin conductivity tensor since the direct product between the irreducible representations of $(Q_x, Q_y, Q_z)$ [or $(G_u, G_v, G_{yz}, G_{zx}, G_{xy})$] and the N\'eel vector belongs to the totally symmetric representation in the magnetic point groups.

We have used the notations of the electric dipole $(Q_x, Q_y, Q_z)$ and the electric toroidal quadrupole $(G_u, G_v, G_{yz}, G_{zx}, G_{xy})$ to represent symbolically any electronic degree of freedom in terms of the polar-vector and second-rank axial-tensor quantities in the Hamiltonian matrix in a broader sense. 
For example, the electric toroidal quadrupole as the bond degree of freedom in the Hamiltonian has been discussed in the spin-orbit-coupled metal Cd$_2$Re$_2$O$_7$~\cite{Hayami_PhysRevLett.122.147602}.
Thus, the symbolic notation of $(Q_x, Q_y, Q_z)$ and $(G_u, G_v, G_{yz}, G_{zx}, G_{xy})$ is applied to both metals and insulators.

\begin{table*}
\caption{
Classification of electric dipole (ED), $(Q_x, Q_y, Q_z)$, and electric toroidal quadrupole (ETQ), $(G_u, G_v, G_{yz}, G_{zx}, G_{xy})$, under cubic, tetragonal, orthorhombic, and monoclinic crystals. 
We take the $x$ axis as the $C_{2}'$ rotation axis.
}
\label{tab_multipoles_table1}
\centering
\begin{tabular}{ll|ccccc|ccccccc|ccc|ccc} \hline\hline
ED & ETQ &
$O_{\rm h}$ & $O$ & $T_{\rm d}$ & $T_{\rm h}$ & $T$ &
$D_{\rm 4h}$ & $D_{4}$ & $C_{\rm 4h}$ & $D_{\rm 2d}$ & $C_{\rm 4v}$ & $C_{4}$ & $S_{4}$ &
$D_{\rm 2h}$ & $D_{2}$ & $C_{\rm 2v}$ & $C_{\rm 2h}$ & $C_{\rm 2}$ & $C_{\rm s}$\\ \hline
--- & $G_{u}$ &
E$_{u}$ & E & E & E$_{u}$ & E &
A$_{1u}$ & A$_{1}$ & A$_{u}$ & B$_{1}$ & A$_{2}$ & A & B &
A$_{u}$ & A & A$_{2}$ & A$_{u}$ & A & A$''$ \\
--- & $G_{v}$ &
& & & & &
B$_{1u}$ & B$_{1}$ & B$_{u}$ & A$_{1}$ & B$_{2}$ & B & A &
A$_{u}$ & A & A$_{2}$ & A$_{u}$ & A & A$''$ \\
$Q_{x}$ & --- &
T$_{1u}$ & T$_{1}$ & T$_{2}$ & T$_{u}$ & T &
E$_{u}$ & E & E$_{u}$ & E & E & E & E &
B$_{3u}$ & B$_{3}$ & B$_{1}$ & B$_{u}$ & B & A$'$ \\
$Q_{y}$ & --- & 
& & & & &
& & & & & & &
B$_{2u}$ & B$_{2}$ & B$_{2}$ & B$_{u}$ & B & A$'$ \\
$Q_{z}$ & --- & 
& & & & &
A$_{2u}$ & A$_{2}$ & A$_{u}$ & B$_{2}$ & A$_{1}$ & A & B &
B$_{1u}$ & B$_{1}$ & A$_{1}$ & A$_{u}$ & A & A$''$ \\
--- & $G_{yz}$ & 
T$_{2u}$ & T$_{2}$ & T$_{1}$ & T$_{u}$ & T &
E$_{u}$ & E & E$_{u}$ & E & E & E & E &
B$_{3u}$ & B$_{3}$ & B$_{1}$ & B$_{u}$ & B & A$'$ \\
---- & $G_{zx}$ & 
& & & & &
& & & & & & &
B$_{2u}$ & B$_{2}$ & B$_{2}$ & B$_{u}$ & B & A$'$ \\
--- & $G_{xy}$ & 
& & & & &
B$_{2u}$ & B$_{2}$ & B$_{u}$ & A$_{2}$ & B$_{1}$ & B & A &
B$_{1u}$ & B$_{1}$ & A$_{1}$ & A$_{u}$ & A & A$''$ \\
\hline\hline
\end{tabular}
\end{table*}

\begin{table*}
\caption{
Classification of electric dipole (ED), $(Q_x, Q_y, Q_z)$, and electric toroidal quadrupole (ETQ), $(G_u, G_v, G_{yz}, G_{zx}, G_{xy})$, under hexagonal and trigonal crystals.  
We take the $x$ axis as the $C_{2}'$ rotation axis.
For $C_{3 {\rm v}}$, we take the $zx$ plane as the $\sigma_v$ mirror plane.
$C_{\rm 3i}=S_{6}$. 
The basis functions are taken as $(x,y)\to x\pm iy$ for $C_{\rm 6h}$, $C_{6}$, $C_{\rm 3h}$, and $C_{\rm 3i}$. 
}
\label{tab_multipoles_table2}
\centering
\begin{tabular}{ll|ccccccc|ccccc} \hline\hline
ED & ETQ & 
$D_{\rm 6h}$ & $D_{6}$ & $C_{\rm 6h}$ & $C_{\rm 6v}$ & $C_{6}$ & $D_{\rm 3h}$ & $C_{\rm 3h}$ &
$D_{\rm 3d}$ & $D_{3}$ & $C_{\rm 3v}$ & $C_{\rm 3i}$ & $C_{3}$ \\ \hline
--- & $G_{u}$ & 
A$_{1u}$ & A$_{1}$ & A$_{u}$ & A$_{2}$ & A & A$_{1}''$ & A$''$ &
A$_{1u}$ & A$_{1}$ & A$_{2}$ & A$_{u}$ & A
\\
$Q_{z}$ & --- & 
A$_{2u}$ & A$_{2}$ & A$_{u}$ & A$_{1}$ & A & A$_{2}''$ & A$''$ &
A$_{2u}$ & A$_{2}$ & A$_{1}$ & A$_{u}$ & A
\\
$Q_{x}$ & $G_{zx}$ & 
E$_{1u}$ & E$_{1}$ & E$_{1u}$ & E$_{1}$ & E$_{1}$ & E$'$ & E$'$ &
E$_{u}$ & E & E & E$_{u}$ & E
\\
$Q_{y}$ & $G_{yz}$ & 
& & & & & & &
& & & &
\\
---  & $G_{v}$ & 
E$_{2u}$ & E$_{2}$ & E$_{2u}$ & E$_{2}$ & E$_{2}$ & E$''$ & E$''$ &
E$_{u}$ & E & E & E$_{u}$ & E
\\
--- & $G_{xy}$ & 
& & & & & & &
& & & &
\\
\hline\hline
\end{tabular}
\end{table*}

\begin{acknowledgments}
This research was supported by JSPS KAKENHI Grants Numbers JP19K03752, JP19H01834, JP21H01037, JP22H04468, JP22H00101, JP22H01183, and by JST PREST (JPMJPR20L8). 
Parts of the numerical calculations were performed in the supercomputing systems in ISSP, the University of Tokyo.
\end{acknowledgments}

\bibliographystyle{apsrev}
\bibliography{ref}

\end{document}